\documentclass[acmsmall]{acmart}

\usepackage{enumitem}
\usepackage{subfig}
\usepackage{amsmath}
\usepackage{pdfrender}

\usepackage{amsthm}
\usepackage[textsize=scriptsize]{todonotes}
\usepackage{booktabs,caption}
\usepackage[flushleft]{threeparttable}
\usepackage{booktabs}
\newcommand{\genuinevideoset}{$\boldsymbol{V_o}$}
\newcommand{\collusivelikesvideoset}{$\boldsymbol{V_l}$}
\newcommand{\collusivecommentsvideoset}{$\boldsymbol{V_c}$}
\newcommand{\collusivechannelsset}{$\boldsymbol{V_s}$}

\newcommand{\collate}{\texttt{CollATe}}

\newcommand{\inv}{^{\raisebox{.2ex}{$\scriptscriptstyle-1$}}}
\theoremstyle{definition}
\newtheoremstyle{observation}
  {0.8em}
  {\topsep}
  {}
  {.5em}
  {\scshape}
  {:}
  {.5em}
  {\thmname{#1}\thmnumber{ #2}\thmnote{ (#3)}}
\theoremstyle{observation}
\newtheorem{observation}{Observation}

\AtBeginDocument{%
  \providecommand\BibTeX{{%
    \normalfont B\kern-0.5em{\scshape i\kern-0.25em b}\kern-0.8em\TeX}}}

\setcopyright{acmcopyright}
\copyrightyear{2021}
\acmYear{2021}
\acmDOI{10.1145/1122445.1122456}

\acmJournal{TIST}
\acmVolume{37}
\acmNumber{4}
\acmArticle{111}
\acmMonth{8}

\begin{document}

\title{Detecting and Analyzing Collusive Entities on YouTube}

\author{Hridoy Sankar Dutta$^\ast$}
\affiliation{%
  \institution{IIIT-Delhi}
  \streetaddress{Okhla Industrial Estate, Phase III}
  \city{New Delhi}
  \state{Delhi}
  \country{India}
  \postcode{110020}
}

\author{Mayank Jobanputra$^\ast$}
\thanks{$^\ast$ Both authors contributed equally to the paper}
\affiliation{%
  \institution{IIIT-Delhi}
  \streetaddress{Okhla Industrial Estate, Phase III}
  \city{New Delhi}
  \state{Delhi}
   \country{India}
  \postcode{110020}}

\author{Himani Negi$^\dagger$}
\thanks{$^\dagger$ The work was done when Himani was an intern at IIIT Delhi.}
\affiliation{%
  \institution{BVCOE, New Delhi}
  \city{New Delhi}
  \country{India}
}

\author{Tanmoy Chakraborty}
\affiliation{%
 \institution{IIIT-Delhi}
 \streetaddress{Okhla Industrial Estate, Phase III}
  \city{New Delhi}
\state{Delhi}
\postcode{110020}
 \country{India}}

\renewcommand{\shortauthors}{Dutta et al.}


\begin{abstract}

  YouTube sells advertisements on the posted videos, which in turn enables the content creators to monetize their videos. As an unintended consequence, this has proliferated various illegal activities such as artificially boosting of views, likes, comments, and subscriptions. We refer to such \textit{videos} (gaining likes and comments artificially) and \textit{channels} (gaining subscriptions artificially) as ``collusive entities''. Detecting such collusive entities is an important yet challenging task. Existing solutions mostly deal with the problem of spotting fake views, spam comments, fake content, etc., and oftentimes ignore how such fake activities emerge via collusion. Here, we collect a large dataset consisting of two types of collusive entities on YouTube -- {\em videos} submitted to gain collusive likes and comment requests, and {\em channels} submitted to gain collusive subscriptions. 
  
  We begin by providing an in-depth analysis of collusive entities on YouTube fostered by various {\em blackmarket services}. 
  Following this, we propose models to detect three types of collusive YouTube entities -- videos seeking collusive likes, channels seeking collusive subscriptions, and videos seeking collusive comments. The third type of entity is associated with temporal information. 
  To detect videos and channels for collusive likes and subscriptions respectively, we utilize one-class classifiers trained on our curated collusive entities and a set of novel features. The SVM-based model shows significant performance with a true positive rate of $0.911$ and $0.910$ for detecting collusive videos and collusive channels respectively. To detect videos seeking collusive comments, we propose \collate, a novel end-to-end neural architecture that leverages time-series information of posted comments along with static metadata of videos.  \collate\ is composed of three components -- metadata feature extractor (which derives metadata-based features from videos), anomaly feature extractor (which utilizes the  time-series data to detect sudden changes in the commenting activity), and comment feature extractor (which utilizes the text of the comments posted during collusion and computes a similarity score between the comments). 
  Extensive experiments show the effectiveness of \collate~(with a true positive rate of $0.905$) over the baselines.
\end{abstract}



\maketitle

\section{Introduction}
Online media is increasingly becoming the most effective medium for sharing ideas, thoughts, and information. The primary reasons behind the large user engagement are the ease of accessibility and ever-evolving attractive sharing facility. The content shared in online media usually includes personal data, documents, photos, and videos. Generally, videos are used to convey meaningful information in a shorter duration, providing unparalleled advantages to content consumers. YouTube, an online video-sharing platform,  allows users to upload, share, or live-stream videos on the Internet. Free service, measurable analytics, availability of multiple genres, and access to broad audiences have made YouTube the most popular platform for both content creators and content consumers. The popularity of a YouTube video is generally determined by the number of likes or comments it receives over a period of time. Similarly, the popularity of a YouTube channel is measured by the number of subscribers. The natural way to gain traffic to a channel/video and make it noticeable is a time-consuming process for content creators. Apart from creating attractive content, oftentimes, the content creators become desperate to figure out effective shortcuts to gain quick popularity of their content and channels. This may lead them to choose unethical and inorganic ways with the aid of blackmarket services to gain popularity within a short duration. The primary objective of an artificial boosting mechanism is to convince the YouTube ranking algorithm to prefer a video/channel over its competitors.

In recent years, YouTube has started a \textit{YouTube Partner Programme}\footnote{https://support.google.com/youtube/answer/72851?hl=en} where content creators can make money from the advertisements and other revenue streams. However, the channel has to meet minimum two requirements -- \textit{at least $1,000$ subscribers} and \textit{at least $4,000$ hours of watch time within the past 12 months}. Furthermore, content creators who are new to YouTube and have managed to attract only a small audience do not get significant engagement despite having fantastic content. It leads them to invest to the blackmarket services to reach out to more audiences and increase their revenues in a faster and effective way. The blackmarket-driven artificial engagement is strictly against the YouTube Terms of Service
and may lead to a permanent ban on the user accounts. To counter this, blackmarket services provide their facilities in such a way that collusive entities (channels and videos) are evaded from being detected by the in-house fake detection algorithms deployed by YouTube. Interestingly, in our dataset, we found $7$ such collusive channels, which are marked as \textit{verified} by YouTube! This clearly shows that YouTube is unable to detect such fraudulent entities, thereby creating an inadequate social space for the entire YouTube population.

Throughout the paper, we use the following nomenclature. A YouTube video is said to be  \textit{collusive} if likes or comments of the video are artificially inflated with the help of blackmarket services. Similarly, a collusive YouTube channel is the one which receives artificial subscriptions from blackmarket services. We use \textit{collusive entities} to refer to both collusive videos and channels.

The current work provides a large-scale investigation and detection of collusive entities on YouTube. We create a unique dataset of collusive entities collected from YouLikeHits (a credit-based freemium blackmarket service). 
\textit{This, to our knowledge, is the first labeled dataset of YouTube collusive entities}. We start our analysis by examining videos submitted to blackmarket services for collusive appraisals (likes and comments) based on two perspectives -- propagation dynamics and video metadata. We then analyze the collusive channels based on location, channel metadata, and network properties. In the collusive channel network, the structural properties of the giant component show that it is a small-world. Further, we are interested in detecting whether a new entity is collusive or not. Since we collected the collusive data directly from the blackmarket websites, we are sure that the collected data does not have any noisy labels. In such cases, instead of collecting a large dataset representing the entire YouTube population, we rather focus on designing sophisticated methods that can leverage the characteristics of entities in one class (collusive in our case) and learn their representations for the prediction task.  
Note that we were unable to extract any temporal information related to like and subscription activities for videos and channels respectively, due to the restrictions and limitations of the YouTube API.

We propose three models to detect three types of collusive entities. The first and the second models utilize one-class classifiers trained only on the collusive entities using video metadata and channel features to detect collusive videos and channels respectively. The third model attempts to detect videos seeking collusive comments. In addition to the video metadata, it uses textual and temporal information of each comment. The temporal information indicates the aggressive patterns of blackmarket users (as these users aggressively post comments on collusive YouTube videos in order to gain credits). In addition to this, the semantic representation of each comment enables us to learn the similarity pattern of these users in posting comments and also to evade the existing fake detection strategies. During the collection of this dataset, we realized that many videos have a limited number of comments despite being posted for collusive comments to the blackmarket services. We discarded such videos as the comments were possibly deleted, which might result in noisy information in the labeled data. Finally, we end up collecting a relatively smaller dataset of videos seeking collusive comments compared to those videos/channels seeking collusive likes/subscriptions. This further raises another challenge of learning generalized and robust representation. Thus, by incorporating these properties of blackmarket users and taking into account the issues mentioned above, we propose \collate, a denoising autoencoder model to detect videos submitted in blackmarket services for collusive comments. \collate\ consists of three components -- metadata feature extractor, anomaly feature extractor, and comment feature extractor to learn feature representation of videos. 

The first and second models achieve significant accuracy with a true positive rate of $0.911$ and $0.910$ respectively. The third model, \collate\ achieves a true positive rate of $0.905$, outperforming seven baselines. Altogether this work sheds light on how collusion happens on YouTube and focuses on the detection of collusive entities to make YouTube an adequate video sharing platform for the content creators and content consumers. We believe this could be used to curb the adverse effects of collusion in gaining artificial social growth. We summarize the contributions of this work as follows:
\begin{itemize}[leftmargin=*]
    \item To the best of our knowledge, we are the first to investigate YouTube videos and channels submitted to blackmarket services for collusive appraisals. 
    
    \item We prepare four unique datasets of collusive entities on YouTube: (i) videos submitted to YouTube for collusive likes, (ii) videos submitted to YouTube for collusive comments, (iii) channels submitted to YouTube for collusive subscriptions, and (iv) the network of collusive YouTube channels. 
    The dataset is attained using YouTube API and custom-designed scrapers that can be easily extended and used to collect large YouTube data in an effective way. We believe that these four datasets would help researchers analyze blackmarket-driven collusive activities happening on YouTube and develop tools to detect them.

    \item We analyze the YouTube videos for collusive likes and comments from two perspectives -- propagation dynamics and video metadata. We also analyze collusive YouTube channels based on their location, channel metadata, and network properties. The giant component of the collusive channel network turns out to be a small-world. 
    
    \item We utilize one-class classification models to detect videos and channels submitted to blackmarkets for collusive likes and subscriptions. We also propose \collate, an end-to-end neural framework to detect YouTube videos seeking collusive comments.

    
\end{itemize}

\textbf{\underline{Reproducibility:}} Codes and datasets are available at the following link: \url{https://github.com/LCS2-IIITD/CollATe}.

\textbf{\underline{Organization of the paper:}} The remainder of the paper is organized as follows. Section \ref{sec:rw} discusses related work. Section \ref{sec:bg} presents a detailed study of the blackmarket services and motivates the research problem. Section \ref{sec:data} discusses the data collection strategy. Section \ref{sec:anatomy} analyzes collusive YouTube videos and channels. Section \ref{sec:automatic_detection} introduces the proposed frameworks to detect collusive entities. The experimental results are presented in Section \ref{sec:result}. Section \ref{sec:implications} shows the important implications of the collusive entities detected using our models. Section \ref{sec:conclusion} concludes the paper with future directions.

\section{Related Work}\label{sec:rw}
We divide the relevant related work into two parts: (i) fraud/spam detection in online media, and (ii) studies on blackmarket services. 

\subsection{Fraud/spam detection in online media}
Plenty of studies focused on the detection of fraudulent activities in a wide range of platforms. In recent years, fraudulent activities are most common in major online media sites such as Facebook \cite{de2014paying,badri2016uncovering,ikram2017measuring}, Instagram \cite{sen2018worth,zhang2017instagram} and Twitter \cite{egele2015towards,gupta2013faking,stringhini2015evilcohort,ghosh2012understanding,thomas2011suspended,wang2010don,nilizadeh2017poised,egele2013compa}. \citet{kumar2018false} presented a review on existing fake and fraud detection strategies in online media platforms. A number of fake comment detection strategies \cite{alsaleh2015combating,uysal2018feature} on YouTube were also proposed in recent years. Most of these studies rely solely on the textual data of the comment. Nevertheless, there is no prior work that considers the temporal properties of comments, which in the case of collusion, is the most important factor, the reason being collusive users perform appraisal operations aggressively in order to gain credits rapidly. \citet{li2016world} proposed \texttt{LEAS}, an algorithm to detect fake engagements in video sharing platforms using a temporal engagement graph between users and video objects. \citet{marciel2016understanding} proposed a set of tools to detect view fraud in online video portals. \citet{chen2015analysis} investigated the problem of fake views caused by robots in video sharing platforms. 

In the field of fraud detection in online advertising, \citet{metwally2007detectives} proposed an advertising network model to discover coalitions between pairs of fraudsters in e-commerce platforms. \citet{dave2012measuring} designed an automated approach for ad networks to detect click-spam attacks. \citet{hussain2018analyzing} analyzed disinformation and crowd manipulation tactics on YouTube by analyzing video metadata. \citet{alassad2019examining} examined intensive groups among YouTube commenter networks by constructing a two-level optimization problem for maximizing local degree centrality and global modularity measures. \citet{faddoul2020longitudinal} conducted a longitudinal analysis of the promotion of conspiracy videos on YouTube. \citet{yang2017fake} studied the problem of injection attacks on the recommendation systems (fake co-visitations) of YouTube.  A number of studies have been conducted on detecting spam on video streaming platforms. \citet{alberto2015tubespam} proposed \texttt{Tubespam}, a novel classification model for comment spam filtering on YouTube. \citet{uysal2018feature} studied the performance of five state-of-the-art text feature selection methods for spam filtering on YouTube using Naive Bayes and Decision Tree. \citet{yusof2017detecting} detected video spammers on YouTube based on the EdgeRank algorithm to decide which post/stories should appear in each user's news feed. \citet{chowdury2013data} proposed a spam detection system for YouTube using a set of spam-related attributes from videos. 
\citet{sureka2011mining} detected forum spammers on YouTube based on the mining comment activity log of a user and extracting patterns indicating spam behavior. \citet{aiyar2018n} detected spam comments on YouTube by showing the effectiveness of using character n-grams instead of word n-grams to improve the accuracy of the classification model. 

More recently, YouTube has become the most important tool for live streamers. In our dataset, we found around $30\%$ of the collusive channels involved in streaming live videos. \citet{zhang2018crowdsourcing,zhang2018end} discussed the detection of copyright infringement on YouTube live videos. \citet{zhang2018crowdsourcing} developed a crowd-sourced based copyright infringement detection (CCID) scheme from live chat messages on YouTube and Twitch to identify original copyright content from the owner. \citet{zhang2018end} presented an end-to-end supervised detection framework to combat copyright infringement in live video streams using the live chat messages from the audiences. However, these methods tend to combat the problem of fake, fraud, and spam detection in video-sharing platforms but are not applicable for collusive entity detection.

\subsection{Studies on blackmarket services}
Despite the fact that a plethora of studies exist on detecting fraud/spam activities in online media, there has been relatively less work on investigating blackmarket services providing collusive appraisals. \citet{acker2018data} focused on how manipulators create disinformation by fake engagement activities on YouTube. \citet{keller2018flourishing} provided a broad overview of the blackmarket services providing fake YouTube views. The authors reported that one of the blackmarket services, named Devumi had earned more than $\$1.2$ million in around $3$ years of service by selling $196$ million views.  

Studies that are more related to the current work include \citet{shah2017many}, \citet{dutta2018retweet}, and \citet{chetan2019corerank}, which investigated the problem of blackmarket-based collusive activities in Twitter. \citet{shah2017many} studied multiple types of blackmarket link fraud behaviors in Twitter by analyzing the connectivity patterns of fake followers via the egonet and boomerang networks.  \citet{dutta2018retweet} proposed \texttt{SCoRe}, a supervised method to detect collusive retweeters affiliated to blackmarket services in Twitter. The authors also showed the differences between fake  and collusive activities based on the synchronicity of retweet behaviors. \citet{dutta2019blackmarket} extended their previous work \cite{dutta2018retweet} to show the differences between the working principles of premium and freemium blackmarket services. \citet{chetan2019corerank} proposed \texttt{CoReRank}, an unsupervised method to detect collusive users and suspicious tweets by leveraging the user's retweeting and quoting patterns. \citet{dutta2020hawkeseye} proposed \texttt{HawkesEye}, a framework to detect fake retweeters using Hawkes process and topic modeling on tweets. \citet{arora2019multitask} proposed a multi-task learning approach to detect tweets submitted to freemium blackmarket services. \citet{dhawan2019spotting} proposed \texttt{DeFrauder}, an unsupervised method to spot online fraudulent collusive groups in review websites. \citet{wang2011review} proposed a review graph model to detect spammers in online review stores. \citet{zhu2016new} proposed an automated approach to detect collusion behavior in online question-answering systems. Other studies  identified fake followers on Twitter \cite{castellini2017fake,cresci2015fame,aggarwal2014followers,stringhini2013follow,dutta2021decife}. \citet{mehrotra2016detection} and \citet{jiang2014catchsync} are some of those who used network-centric properties to detect fake followers. \texttt{Fake Follower Check\footnote{\url{https://tinyurl.com/y29t3uuu}}} is one such tool to detect fake followers based on profile-centric and behavioral features of Twitter users. We encourage the reader to go through \cite{dutta2020blackmarket} for a detailed survey of collusion in online media platforms.

To the best of our knowledge, there has been no major attempt to detect collusive entities on video sharing platforms such as YouTube. Our current effort provides a deeper understanding of the collusive activities on YouTube and focuses on designing automated approaches to detect these activities. 

\section{Background and Motivation}\label{sec:bg}
\subsection{Blackmarket services}
Blackmarket services help online media users in gaining appraisals inorganically for their content. They offer services related to online social networks (Facebook: like, follow, share; Instagram: like, follow; Twitter: retweet, like, follow), recruitment platforms (LinkedIn: endorsement, recommendation, connection), video sharing platforms (YouTube: video views, video comments, channel subscription; Vimeo: video views, video comments), etc. The inorganic appraisals help in artificial boosting of online media content, thereby creating an inadequate social space. Online media entities such as big companies, advertising firms seeking active participation in their promotional campaigns target these websites to expedite their reach to their target audiences.

The blackmarket services are divided into two types based on the mode of service \cite{malhotra2015long}:
\begin{itemize}
    \item \textbf{Premium services:} These services charge customers for the facilities they provide. Customers have to register themselves and opt for one of their plans to gain appraisals.
\item \textbf{Freemium services:} These services are free of cost and work like a barter system. The primary goal of freemium services is to let  their customers  familiarize with free services and convince them to subscribe to the premium plans. Most of the freemium services are \textit{credit-based}, where each customer receives virtual credits by appraising the content of other customers.
\end{itemize}

In this work, we focus our analysis on \textit{YouLikeHits}\footnote{\url{https://www.youlikehits.com/}}, a credit-based freemium service which helps content creators on YouTube to boost their subscribers, views, comments and likes.

\subsection{Collusion on YouTube platform}
YouTube is a video sharing platform where users can upload videos by creating YouTube channels. The platform is free of cost and is operated by two types of users: \\
(i) \textit{Content creators:} Users who upload videos to their channel. \\
(ii) \textit{Content consumers:} Users who watch videos, interact with videos in the form of likes/comments, or subscribe to channels.

When a content creator uploads a video, content consumer can perform the following actions --\textit{like} the video, \textit{dislike} the video, \textit{comment} on the video, \textit{share} the video, \textit{save} the video to wishlist, and \textit{subscribe} the content creator's channel. The popularity of a YouTube video is measured by the number of appraisals it receives from the content consumers. Thus the content creators need to ensure that their content receives high appraisals from the consumers. Moreover, with the advent of the concept of \textit{Monetization on YouTube\footnote{\url{https://www.youtube.com/account_monetization?nv=1}}}, content creators have begun to attract audiences by uploading videos aimed towards specific genres such as teaching, entertainment, business, etc. This may further motivate them to choose artificial ways to gain quick popularity in their content. Currently, the earning potential of a channel/video is solely driven by the number of subscribers/views. When a video is posted on YouTube, it is shared with the YouTube community of similar channels as recommendations. With millions of videos posted every second, a majority of the videos go unnoticed to the target audience. The organic way of gaining appraisals is a tedious task, which leads content creators to opt for an alternative way by means of blackmarket services.

\paragraph{{\bf How collusion happens on YouTube?}}
Collusion on YouTube happens when a video or a channel is posted in blackmarket services for appraisals. In this work, we refer to a user who submits the content to the blackmarket service as a \textit{collusive user}. In a freemium blackmarket service, collusive users receive credit points upon performing appraisals on the content of other collusive users. In the case of YouTube, the majority of the blackmarket services request the collusive users to submit only the video or channel URL. Collusive users can contribute to the artificial boosting of YouTube videos and channels in several ways: (i) viewing other videos, (ii) posting a comment on videos\footnote{Note that blackmarket services may ban a collusive user upon identifying a spammy comment posted by the user.}, (iii) posting likes on a video, and (iv) subscribing to a channel.

\begin{figure}[!t]
    \centering
    {\includegraphics[width=0.62\linewidth]{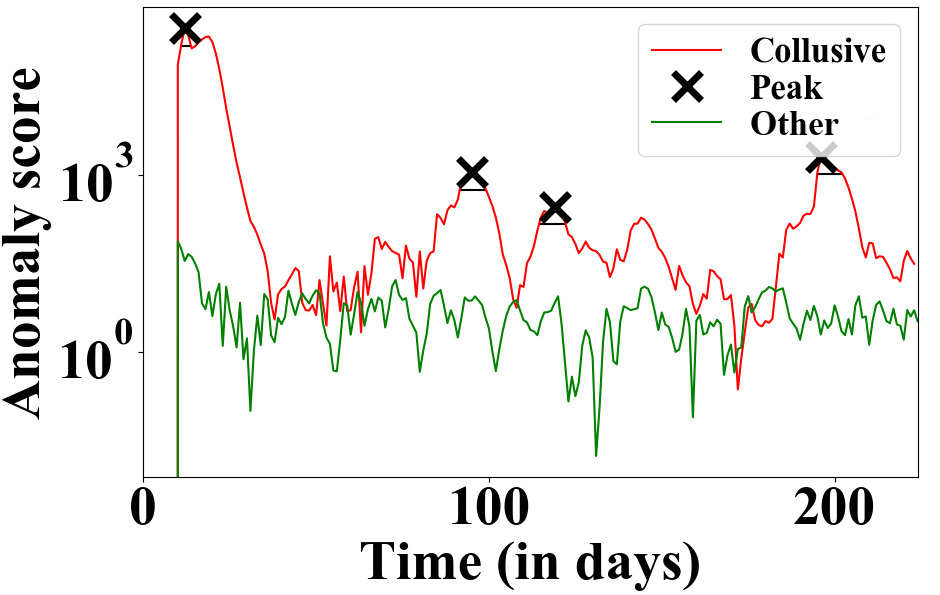} }
    \caption{An example of anomalous pattern in videos detected by \collate~for collusive (in red) and other videos (in green). The x-axis displays the time span (in days)
 starting from when the first comment was posted and y-axis determines an anomaly score calculated by \collate. Here, the anomaly score is the Mahalanobis distance computed using Equation \ref{eq:anomaly} as mentioned in Section \ref{sec:peak_detection}. The peak width  (horizontal black line) corresponding to every peak indicates the duration of the peak. }
    \label{fig:motivation}%
\end{figure}

 Fig. \ref{fig:motivation} shows the peaks observed in collusive videos (in red) for collusive comment appraisals detected using our proposed approach, \collate. However, we do not see any such peaks in other random videos (more details can be found in Section \ref{sec:video_detector}). The reason is that the collusive users tend to perform collusive appraisals aggressively to gain credit points, which they can use later to add new content. This aggressive nature results in peaks in the inter-arrival time of appraisals of the collusive entities; however, normal users do not exhibit such behavior. 

\paragraph{{\bf Why only YouTube?}}
In this article, our focus is to detect collusive entities only on YouTube. The primary reason of conducting this study in one platform is due to the challenge in collecting a large dataset of videos and channels hosted on other platforms and submitted to blackmarket services for collusive appraisals. While conducting this study, we checked a few freemium blackmarket services which provide collusive appraisals to video-sharing platforms such as Twitch and Vimeo. During the collection of information from the blackmarket services, we observed that a very less data is available for these platforms. Additionally, unlike YouTube, the APIs of these platforms have certain limitations and do not provide enough public information which can be used for detailed analysis of the collected entities.

\paragraph{{\bf Challenges in collusive entity detection.}}
Detecting collusive entities is a difficult task \cite{dutta2018retweet,dutta2019blackmarket}. There exist blackmarket services which have created intelligent mechanisms to produce collusive appraisals. We list down the unique challenges of the collusive entity detection task below:
\begin{enumerate}[leftmargin=*]
	\item Collusive accounts are not full-time bots, spam-accounts or fake accounts used only for appraisal. Thus, existing studies focusing on bot, spam and fake account detection are not able to capture their behavior as shown in \cite{dutta2018retweet,dutta2019blackmarket}.
	\item Collusive accounts do not show a completely genuine behavior. Some of their appraised content would be endorsed not because of their interest in the content, but merely to comply with the barter system in blackmarket services. This implies that collusive users show an amalgamation of both organic and inorganic behavior -- being genuine users, they organically appraise the content of their interest; being a blackmarket members, they also inorganically  appraise the content posted in blackmarket services.
	\item Limited contextual information is available on short texts present in collusive entities such as comments, replies, etc., which makes it difficult for deep neural networks to generate appropriate representations. 
\end{enumerate}

\section{Dataset description}\label{sec:data}
\subsection{Data collection} \label{sec:data_collection}
The major challenge is to collect a large set of YouTube videos and channels submitted to the blackmarket services for collusive appraisals and a contrasting set of videos. We started our data collection  by designing multiple web scrapers for the following purposes -- (i) scraping data from blackmarket website (YouLikeHits) and (ii) scraping video related data (i.e., description, comments) of YouTube videos. Both the scrapers performed their operation independently. We used Joblib\footnote{https://joblib.readthedocs.io/en/latest/} library to utilize multiple cores in the deployed server. 
We also extensively used YouTube API\footnote{\url{https://developers.google.com/youtube/v3/}} for the following purposes -- (i) extracting subscribers of YouTube channels, and (ii) extracting the exact time at which the comments are posted. 

We collected the information of collusive videos from YouLikeHits, a blackmarket service that provides three types of collusive appraisals -- (i) likes to YouTube videos, (ii) comments to YouTube videos, and (iii) subscriptions to YouTube channels. We queried multiple search engines with keywords such as `free YouTube likes', `free YouTube comments'. Interestingly, apart from links to the websites providing artificial YouTube views and subscriptions, we found a large number of blackmarket websites pointing to other online media platforms such as Twitter, Facebook, Instagram. It indicates the popularity of blackmarket websites to achieve artificial social status in a much rapid way among online media users.  We developed a scraper to retrieve YouTube entities (videos/channels) involved in collusive appraisals on YouLikeHits. Interestingly, we observe that some of the collusive videos are propagated by verified YouTube channels. We found $342$ ({\em resp.} $86$) videos submitted to YouLikeHits for collusive likes ({\em resp.} comments). We also found $7$ verified YouTube channels, which are registered in YouLikeHits for collusive subscriptions. While scraping, we found that only $25\%$ of videos for collusive likes and $4.05\%$ of videos for collusive comments are deleted by YouTube's current fraud detection system. All the above observations show that YouTube is unable to detect these entities effectively using its in-house fraud detection mechanism. These insights further motivated us to design efficient methods to detect YouTube entities that are involved in gaining collusive appraisals with the help of blackmarket services. 

\begin{figure}[!t]
    \centering
    \subfloat[]{{\includegraphics[width=4.6cm]{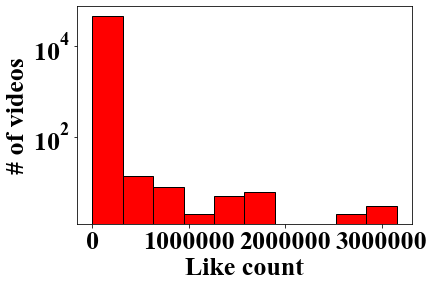} }}%
    \subfloat[]{{\includegraphics[width=4.6cm]{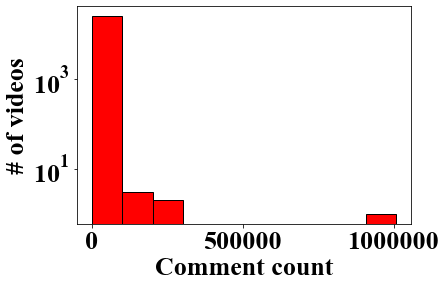} }}%
    \subfloat[]{{\includegraphics[width=4.5cm]{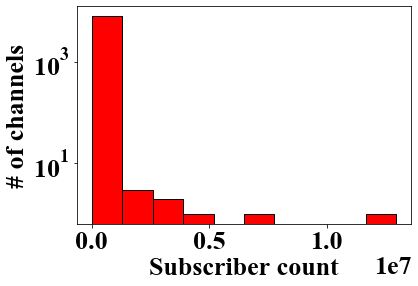} }}%
    \caption{Distribution of (a) likes of YouTube videos, (b) comments of YouTube videos, and (c) subscribers of YouTube channels in our dataset.}%
    \label{fig:dist}%
\end{figure}

We extracted video metadata and comments of all the videos using our custom-designed scrapers and YouTube API. Finally, we divided the dataset into three unique sets  -- \collusivelikesvideoset\ (videos submitted to YouLikeHits for collusive likes), \collusivecommentsvideoset\ (videos submitted to YouLikeHits for collusive comments) and \collusivechannelsset\ (channels submitted to YouLikeHits for collusive subscriptions). The entire data statistics is showed in Table \ref{table:data_statistics} and Fig. \ref{fig:dist}. Note that we did not find any videos which are common across \collusivelikesvideoset\ and \collusivecommentsvideoset. 


\begin{table}[!t]
\caption{Summary of the dataset. Here, the column \textit{\# unique CC} refers to the number of unique content creators of videos submitted for collusive likes and comments. \textit{Entities} refers to \textit{videos} or \textit{channels}; \textit{actions} refers to \textit{like/comment} for videos and \textit{subscription} for channels. }

\label{table:data_statistics}
\begin{center}
\scalebox{0.90}{
\begin{tabular}{|c|c|c|c|c|c|c|c|}
\hline
\centering \textbf{Type} & \textbf{\# entities} & \textbf{\# deleted} & \textbf{\# verified} & \textbf{\# unique CC} & \textbf{Max actions} & \textbf{Min actions} & \textbf{Avg. actions}\\
\hline
\hline \collusivelikesvideoset & $45572$ & $15662$ & $342$ & $28702$ & $3151770$ & $0$ & $1333$\\
\hline \collusivecommentsvideoset & $25106$ & $1060$ & $86$ & $11752$ &  $1008428$ & $0$ & $120$ \\
\hline \collusivechannelsset & $7847$ & $0$ & $7$ & $-$ & $12935205$ & $0$ & $5378$\\
\hline
\end{tabular}}
\end{center}
\end{table}
\subsection{Data privacy}
To our knowledge, this work is the first effort that aims to analyze YouTube videos and channels submitted to blackmarket services for collusive appraisals. We emphasize that we will not release the sensitive information (i.e., video ids and uploader details) when we make the dataset public. The entire data collection process was performed after taking proper Institutional Review Board (IRB) approval from our institute.

\section{Analyzing collusive YouTube entities}
\label{sec:anatomy}
We analyze the collusive videos from two aspects - (i) propagation dynamics, and (ii) video metadata. In the first part of this section, we will present the propagation dynamics of collusive videos using two metrics -- \textit{initial burst} and \textit{lifetimes}. In the second part, we will show the analysis of the collusive YouTube channels based on location, channel metadata and network structure.

\subsection{Videos submitted for collusive comments/likes} \label{sec:features_collusive_likes}
For videos submitted for collusive comments, we extract the video metadata and video comments (full text of each comment and timestamp at which the comment was posted). Note that due to the restrictions of YouTube API, we are unable to provide detailed insights of the videos submitted to blackmarket websites for collusive likes. The API only allows retrieving the total count of likes and dislikes.
\begin{figure}[!t]
    \centering
    \subfloat[]{{\includegraphics[width=6cm]{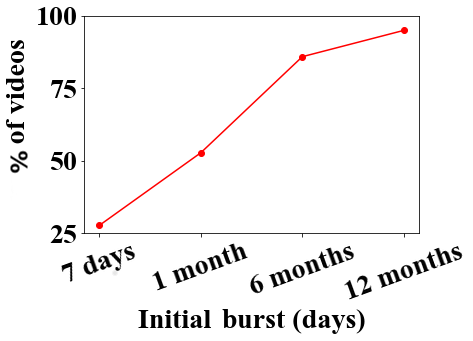} }}%
    \subfloat[]{{\includegraphics[width=6cm]{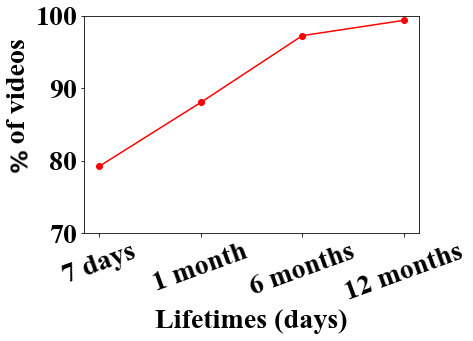} }}%
    \caption{Distribution of temporal properties characterizing propagation dynamics -- (a) initial delays and (b) lifetimes of Youtube collusive videos.}%
    \label{fig:propagation_dynamics}%
\end{figure}
\subsubsection{Propagation dynamics of artificial boosting}
Here we focus on the temporal properties of collusive videos based on two features -- \textit{initial burst} and \textit{lifetimes}. As the timestamps of occurrence of like activities are not available with us due to API restrictions, we perform the analysis only on videos submitted to blackmarkets for collusive comments. \\
(i) \textbf{Initial burst:} We consider the initial burst as the first time when there is a peak in the arrival rate of comments in a video. Section \ref{sec:peak_detection} outlines the peak detection technique. Through this analysis, we characterize how rapidly a video receives collusive appraisals.  Fig. \ref{fig:propagation_dynamics}(a) shows that around $30\%$ of the videos have an initial burst of artificial boosting (first peak) within 7 days, and around $50\%$ of the videos have an initial burst within the first one month from the date of posting. \\
(ii) \textbf{Lifetimes:} Here we consider the lifetime of the collusive activity over a video. We calculate the delay between the burst of the first peak and the fall of the last peak. Fig. \ref{fig:propagation_dynamics}(b) shows that about $80\%$ of the videos have lifetimes within $7$ days. This illustrates how these videos gain rapid attention through blackmarket services.

\subsubsection{Metadata of collusive videos}\label{sec:metadata}
\begin{figure}[!t]
    \centering
    {\includegraphics[width=0.7\linewidth]{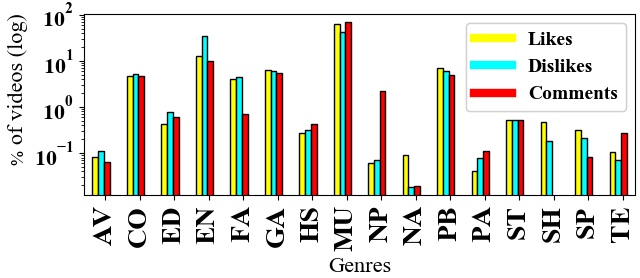} }
    \caption{Genre-wise distribution of likes, dislikes and comments of collusive videos. Full forms of the labels in the x-axis are mentioned in Section \ref{sec:metadata}.}
    \label{fig:genreWiseVideos}%
\end{figure}

\begin{figure}[!t]
    \centering
    \subfloat[]{{\includegraphics[width=6cm]{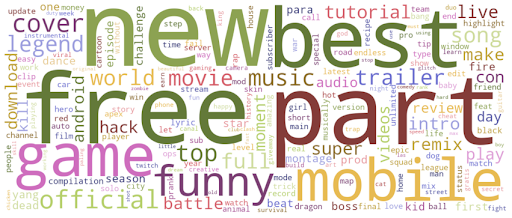} }}%
\subfloat[]{{\includegraphics[width=6cm]{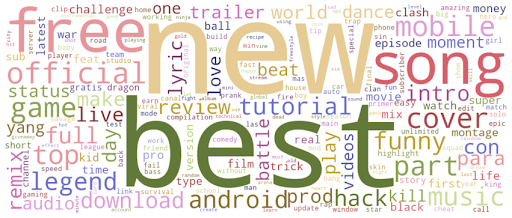} }}%
    \caption{Wordcloud of titles of videos submitted for (a) collusive comments and (b) collusive likes.}%
    \label{fig:wordcloud_video}%
\end{figure}

We now analyze the metadata of collusive videos. We study the genres of the videos and the titles of videos submitted to blackmarket for collusive appraisals. \\
\textbf{(i) Video genres:} 
We observe the distribution of genre of videos submitted to blackmarket services. The common genres on YouTube are -- Gaming (GA), Entertainment (EN), Travel \& Events (TE), Film \& Animation (FA), Music (MU), People \& Blogs (PB), Autos \& Vehicles (AV), Education (ED), Comedy (CO), News \& Politics (NP), Howto \& Style (HS), Science \& Technology (ST), Sports (SP), Pets \& Animals (PA) and Nonprofits \& Activism (NA). We plot the genre-wise distribution of likes, dislikes and comments of collusive videos in Fig. \ref{fig:genreWiseVideos}. The bars for likes and dislikes are drawn from the video metadata for collusive likes, and bars for comments are drawn from the video metadata for collusive comments. As expected, we observe `Music' to be the most popular genre for videos submitted for collusive likes  ($63.05\%$) and comments ($69.81\%$).\\
\noindent\textbf{(ii) Wordcloud of video title:} We show the wordcloud generated from the title of the videos submitted to blackmarkets for collusive likes and comments in Fig. \ref{fig:wordcloud_video}. For clarity, we remove the two-letter words and common stopwords. Here the font size corresponds to the frequency of the text. We clearly observe the presence of similar keywords such as promotional keywords like `free', `best', `top', etc. in both the cases. With the presence of these keywords, it is evident that videos for collusive like/comment appraisals focus on target-specific keywords for quick promotion.    

\noindent \textbf{(iii) Uploader authenticity:} We also study the authenticity of the uploader of videos for collusive comments and  likes. Surprisingly, we observe that  verified users (marked by YouTube) are also involved in gaining collusive appraisals via blackmarket services (see Table \ref{table:data_statistics}).

\subsection{Channels submitted for collusive subscription requests}
\begin{figure}[!t]
    \centering
    \subfloat[]{{\includegraphics[width=6.6cm]{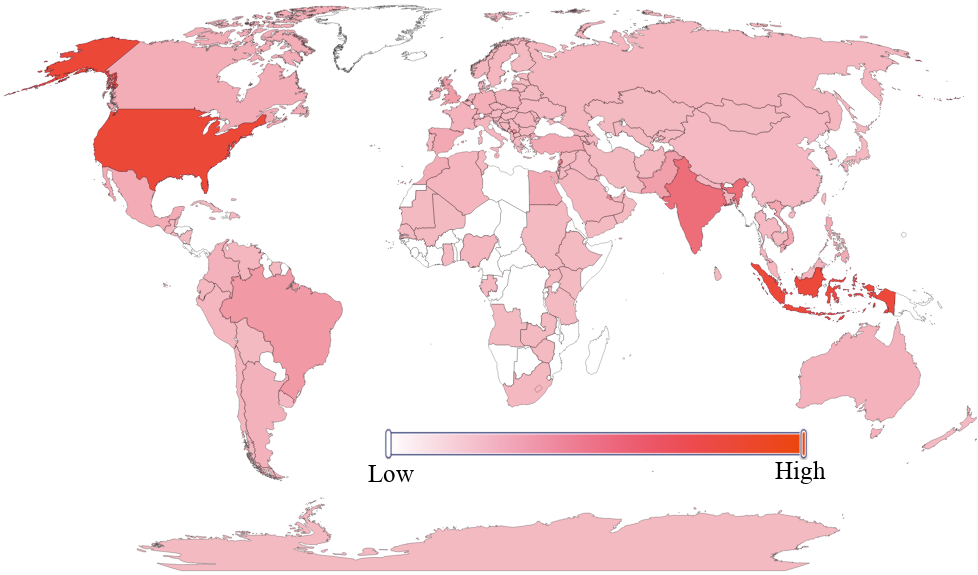} }}%
    \subfloat[]{{\includegraphics[width=4.1cm]{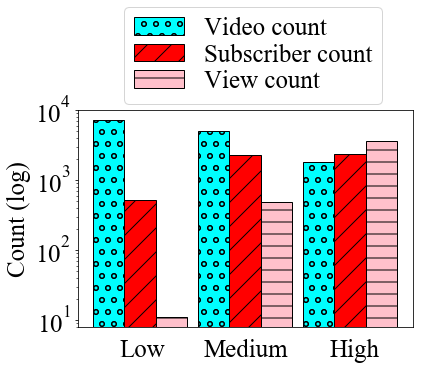} }}%
    \subfloat[]{{\includegraphics[width=3.4cm]{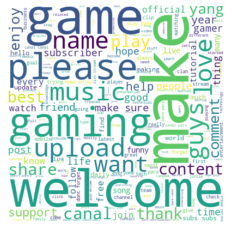} }}%
    \caption{(a) Country-wise distribution of collusive YouTube channels, (b) distribution of YouTube channels based on video, subscriber and views, and (c) wordcloud of channel titles.}%
    \label{fig:channel_dist}%
\end{figure}

In our analysis thus far, we focused only on YouTube video submitted to blackmarket services for collusive likes and comments. In this section, we examine YouTube channels submitted to blackmarket services for collusive subscriptions. \\
\textbf{(i) Country-wise distribution:}
Fig. \ref{fig:channel_dist}(a) shows the world map plot for the country-wise distribution of collusive YouTube channels (in red color), indicating the count of channels submitted to blackmarket services for collusive subscriptions. Out of $7,847$ collusive channels, we find $3,804$ channels with no countries mentioned in the profile. In the remaining set, USA tops the list with $20.11\%$ of collusive YouTube channels followed by Indonesia with $19.58\%$.\\      
%
\textbf{(ii) Metadata of collusive YouTube channels:}
We show the video, view and subscriber count of collusive YouTube channels. For better visualization of the distributions, we create custom range of the counts -- Low ($1 < count < 100$), Medium ($100 <= count <= 1000$) and High ($>1000$). Fig. \ref{fig:channel_dist}(b) shows the distribution of video, subscriber and view count for YouTube channels submitted for collusive subscriptions. We observe that video count and subscriber count are comparable across all ranges; however, there are too many channels with high view count registered in blackmarkets for collusive subscriptions. Fig. \ref{fig:channel_dist}(c) shows the wordcloud aggregated over channel titles. We eliminate common stop words and two-letter word. Here also, the font size corresponds to the frequency of the text. We observe that collusive channels have keywords such as `game', `tutorial', `music', etc., corresponding to personal interests, which help them improve the outreach by connecting with the viewers and subscribers within the same domains.\\
\textbf{(iii) Network observations:}
We study the network structure of collusive channels using social network analysis tools. First, we create an undirected network -- nodes are the collusive YouTube channels, and an edge represents common subscribers among two channels. We use the YouTube API to collect the subscribers of each channel.
Out of $7,847$ channels, we observe only $2,721$ channels with public subscriptions (i.e., end-users can see what channels they are subscribed to). This forms the node-set in the network. We obtain $2,650$ edges among these nodes. We obtain $168$ different connected components from this graph. We consider the final graph to be the maximum connected component (giant component) with $1,320$ nodes and $1,396$ edges for our analysis. We show various network properties (see Table \ref{tab:network_stats}) to better understand the structure of the giant component. 
We notice that a large fraction of nodes ($48.51\%$) belong to the largest connected component, which perhaps indicates the barter system in freemium services. The average shortest path length of the network is $9.037$, which is of the order of the comparable random network of the same size and average degree. Note that the comparable random network is created using the same number of nodes and edges as that of the collusive YouTube channel graph but with random edge connections. The average clustering coefficient of the giant component in the collusive channel network is $0.0023$, which is an order of magnitude higher than that of the comparable random network whose clustering coefficient is $0.0016$. These two structural properties of the giant component indicate that the network of collusive YouTube channels is a small-world \cite{bagler2008analysis}.

\begin{table}[!t]
\caption{Statistics  of collusive YouTube channel network.}
\label{tab:network_stats}
\begin{center}
\begin{tabular}{|c|c|c|c|c|c|}
\hline
\# nodes & \# edges & AD* & Diameter & APL** & Density \\ 
\hline
\hline $1320$ & $1396$ & $2.115$ & $19$ & $9.037$ & $0.0016$  \\
\hline
\end{tabular}
\end{center}
\begin{tablenotes}
    \small \item * - Average degree. ** - Average path length.
\end{tablenotes}
\end{table}


\begin{table}[!htb]
    \begin{minipage}{.5\linewidth}
    
      \caption{Notations and denotations.}\label{tab:notation}
      \centering
        \scalebox{0.7}{\begin{tabular}{l|l}
    \hline
    {\bf Notation} & {\bf Denotation} \\\hline
$TS$ & Time sequence vector of comments\\
$c^{(i)}$ & Cumulative comment count at each timestamp \\
$e$ & Error vector \\
$\mathcal{N}(\mu,\eta)$ & Gaussian distribution using the error vectors \\
$a(c)$ & Anomaly score calculated using Mahalanobis distance \\
$\eta$ & Comment similarity score \\
$P$ & Number of peaks \\
$\textit{W}$ & Number of windows \\
$q$ & Query comment \\
$C$ & Window comments \\
$g(q)$ & Embedding for the query comments \\
$g(C)$ & Embedding for the window comments \\
$x \in X$ & Input data point $x$ in input space $X$\\
$V$ & Intermediate decoded space\\
$Z$ & Intermediate encoding\\
$y \in Y$ & Output label $y$ in labels space $Y$ \\
$\textit{$\hat{x}$}$ & Noisy or corrupted input data point\\
$\tau$ & Encoder present in the autoencoder\\
$\psi$ & Decoder present in the autoencoder\\
\hline
    \end{tabular}}
    \end{minipage}%
    \begin{minipage}{.5\linewidth}
      \centering
        \caption{Abbreviations used throughout the paper.} \label{tab:abbv}
        \scalebox{0.7}{\begin{tabular}{l|l}
    \hline
    {\bf Abbreviation} & {\bf Description} \\\hline
$MFE$ & Metadata Feature Extractor\\
$AFE$ & Anomaly Feature Extractor\\
$CFE$ & Comment Feature Extractor\\
$DAC$ & Denoising Autoencoder Classifier\\
$ARIMA$ & Auto Regressive Integrated Moving Average\\
$WMD$ & Word Mover Distance\\
\hline

        \end{tabular}}
    \end{minipage} 
\end{table}

\section{Detecting collusive entities}
\label{sec:automatic_detection}
In the previous section, we have discussed how collusive videos and channels have an adverse effect on YouTube social space. Therefore, the question we would like to answer is -- \textit{can we automatically detect if an entity on YouTube is submitted for collusive appraisals}?  

In this section, we discuss the methodology for the detection of videos and channels submitted to blackmarket services for collusive likes, comments, and subscriptions. First, we will show how we detect videos for collusive likes, and channels for collusive subscriptions. We pose this problem as a one-class classification problem as the set of genuine videos is unknown to us\footnote{One can argue that YouTube verified channels/videos might constitute the genuine set. We did not consider them because of the following reasons. (i) In our dataset, we found 7 verified channels which were submitted to the blackmarkets. Therefore, we were unsure whether all verified entities are really genuine or not. (ii) Verified channels/videos are usually very popular, receiving huge appraisals from the viewers. Therefore, they may not look like a normal, random YouTube entity and may lead to bias in the dataset.}.
We utilize several one-class classifiers with the proposed sets of static features for the detection.
Secondly, we develop a novel end-to-end framework, named \collate\ that leverages video metadata, anomalous activities and comment similarity for detecting videos for collusive comments. Table \ref{tab:notation}
summarizes the notations with their denotations and Table \ref{tab:abbv} summarizes the abbreviation and description used in the model.

\paragraph{{\bf Reason behind proposing three collusive entity detection tasks.}}
YouTube has three types of appraisals: \textit{likes, comments, subscriptions} for two types of entities: \textit{videos} and \textit{channels}. Likes and comments are for videos, and subscriptions are for channels. In case of collusion, the blackmarket services provide appraisals in such a way that one kind of appraisal is not related to other e.g., if a customer requests for likes on a YouTube video, the video will only receive likes and not any other appraisal. This is also seen in case of Twitter \cite{dutta2019blackmarket} where there are two types of appraisals, \textit{followers} and \textit{retweets}, for two types of entities, \textit{users} and \textit{tweets}, respectively. As each appraisal is independent of other appraisals, we proposed three detection tasks and did not combine them to just one task.

\paragraph{{\bf Reason behind proposing two one-class models and one binary classification model.}}
The reason behind not framing the first two models as a binary classification task is the unavailability of textual and temporal information for these two models. In literature, it has been seen that collusive accounts show a hybrid behavior -- they sometimes behave like a genuine accounts and exhibit organic activities; at the same time, they inorganically appraise the content of other accounts to gain credits from blackmarket services. This made us to hypothesize that collusive accounts tend to have very diverse topics of interest as they appraise content without their genuine interest (textual information) and show sudden changes in the commenting activity (temporal information). In our case, the appraisal present in the third task (i.e., collusive comments for YouTube videos) is an interactive-based appraisal where a user posts a textual comment on a specific time for a video. However, appraisals present in the first two tasks, i.e., collusive likes for YouTube videos and collusive subscriptions for YouTube channels, are tap-based appraisals where a user simply taps on the like/subscribe button. The only way to collect temporal information for these appraisals is using YouTube Live Streaming API which has a quota limitation and makes it infeasible in a real-time scenario where data comes in streaming fashion. Moreover, tap-based appraisals do not contain any textual information. Thus, we proposed the first two models as one-class classification models and the last model as a binary classification model.

\subsection{Features for detecting videos submitted for collusive likes} \label{sec:collusive_like_detection}
In this section, we present novel features to characterize videos submitted to blackmarket services for collusive like requests. In Section \ref{sec:prediction_results}, we show how these features affect the performance of different one-class classification models.

\begin{figure}[!t]
    \centering
    \subfloat[]{{\includegraphics[width=3.6cm]{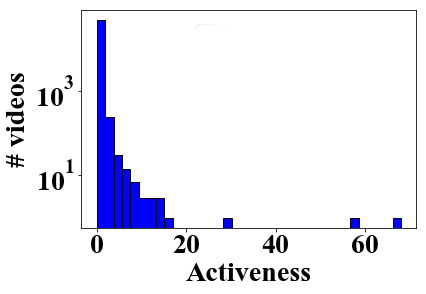} }}%
    \subfloat[]{{\includegraphics[width=3.6cm]{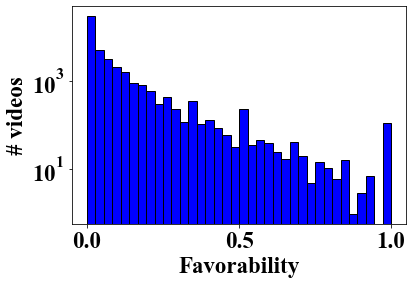} }}%
    \subfloat[]{{\includegraphics[width=3.6cm]{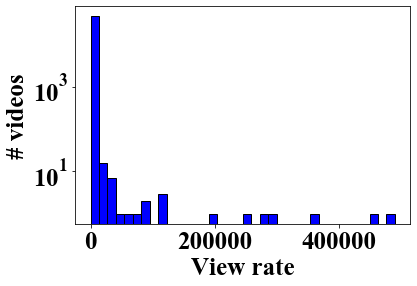} }}%
    \subfloat[]{{\includegraphics[width=3.6cm]{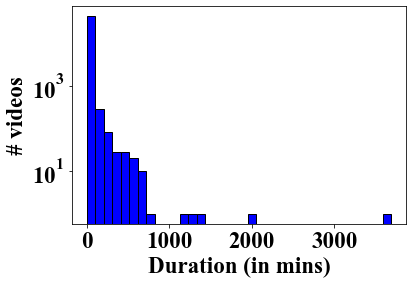} }}%
    \caption{Exploratory analysis of features used to characterise videos submitted for collusive likes.}%
    \label{fig:analysis_collusive_likes}%
\end{figure}

$\bullet$ \textbf{{Activeness ($\alpha$):}} We observe that collusive videos generally receive more likes with a relatively small number of views. We compute the \textit{Activeness} as the ratio of the total number of likes to the total number of views gained by the video. It shows the high engagement of content consumer on the basis of likes and views. Fig. \ref{fig:analysis_collusive_likes}(a) shows the distribution of activeness for collusive videos. 

$\bullet$ \textbf{{Favorability ($\beta$):}} We notice that collusive videos receive less dislikes even for a large number of views\footnote{Note that the number of dislikes is not complementary of the number of likes, as users who view a video may not make any action (like/dislike). Therefore, it is not necessary that the sum of likes and dislikes of a video is the number of views.}. We compute the \textit{Favorability} as the ratio of dislikes to the sum of likes and dislikes gained by the video. It shows the likelihood of likes gained by the video. Fig. \ref{fig:analysis_collusive_likes}(b) shows the distribution of favorability for collusive videos.

$\bullet$ \textbf{{View-rate ($\gamma$):}} The collusive users may focus on increasing the likes of the videos. However, with the growth in likes, view count will automatically increase (the reverse is not true though). This unintended growth in views also helps the content creator to gain complementary collusive views. We compute the \textit{view-rate} as the ratio of the total number of views gained by the video to the total number of days since its submission to YouTubey. Fig. \ref{fig:analysis_collusive_likes}(c) shows the distribution of view-rate for collusive videos.

$\bullet$ \textbf{{Video duration ($\delta$):}}
We observe that YouTube videos for collusive comment requests are fairly short length videos with an average duration of $6$ minutes. We consider the length of the video (in seconds) as one of the features for the collusive entity detection. Fig. \ref{fig:analysis_collusive_likes}(d) shows the distribution of video duration for collusive videos.

The final metadata feature extractor \textit{$v_e$} has the following form:
\begin{equation}
    v_e = (\alpha,\beta,\gamma,\delta)
\end{equation}

\subsection{Features for detecting channels submitted for collusive subscriptions } \label{sec:collusive_channel_detection}
In this section, we present the novel features to characterize channels submitted to blackmarket services for collusive subscriptions. Note that due to YouTube API restrictions, we are unable to collect granular details about the YouTube channels.

$\bullet$ \textbf{{Hidden subscriber count:}} It is the number of subscribers the channel receives which are hidden from its profile. 

$\bullet$ \textbf{{Video count:}} It is the count of the total number of videos present in the channel.

$\bullet$ \textbf{{Subscriber count:}} It is the count of the total number of subscribers the channel has received.

$\bullet$ \textbf{{View count:}} It is the total number of views the channel has received. 

$\bullet$ \textbf{{Comment count:}} It is the total number of comment the channel has received.

\subsection{Detecting videos submitted for collusive comments} \label{sec:collusive_comment_detection}
In this section, we present novel features to characterize videos submitted to blackmarket
services for collusive comments. In Section \ref{sec:prediction_results}, we will show the importance of these features. It is known that not all the comments posted on a video are collusive in nature. This makes it hard to identify collusive comments even for human experts. Our dataset contains rich social information about YouTube videos such as uploader details, video metadata, channel details, raw comments text, and comment timestamp. The goal of our proposed model is to transform the data into useful features and identify the collusive videos.

\begin{figure*}[!t]
    \centering
    {\includegraphics[width=\linewidth]{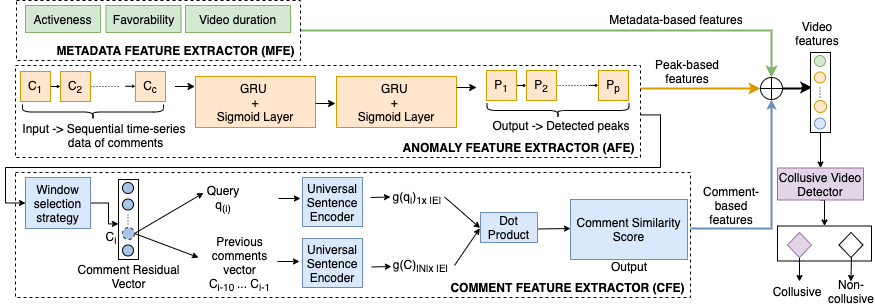} }
    \caption{The architecture of \collate. The green colored network is the metadata feature extractor, the orange colored network represents the anomaly feature extractor, and comment feature extractor is blue colored.}%
    \label{fig:architecture}%
\end{figure*}

\collate\ comprises three components: (i) metadata feature extractor ($v_m$), (ii) anomaly feature extractor ($v_a$), and (iii) comment feature extractor ($v_c$) as shown in Fig. \ref{fig:architecture}. Once $v_m$, $v_a$ and $v_c$ are derived, they are concatenated to create the final video representation. The collusive video detector takes the learned representations as input to predict if a video is collusive. The details of the metadata feature extractor is mentioned in Section \ref{sec:collusive_like_detection} (we use the same metadata feature extractor except view-rate ($\gamma$)). The reason behind not choosing view-rate is that a popular YouTube video is likely to have similar view-rate to the collusive one.  Two other components are mentioned below.

\subsubsection{Anomaly feature extractor}\label{sec:peak_detection}
The goal of the anomaly feature extractor is to detect sudden changes in the commenting activity over a video. Such activities generally continue for a shorter time duration and then stop suddenly. We would like to detect such activities for a given video to derive useful features. The anomaly feature extractor takes sequential time-series data of comments as input. To extract the anomalies (peaks) from the time-series data, we employ Gated Recurrent Units (GRU) as the core module of the anomaly feature extractor. We adapt the idea of stacking recurrent layers for anomaly detection from Malhotra et. al \cite{malhotra2015long}. 

\textbf{Overview:} Consider a time sequence vector $TS = \{ c^{(1)}, c^{(2)} \allowbreak c^{(3)}, \dots  c^{(n)} \}$ where every point $c^{(t)} \in R^m $ in the time sequence 
is a $m$-dimensional vector $\{c^{(t)}_1, c^{(t)}_2, \allowbreak c^{(t)}_3, \dots  c^{(t)}_m \}$, denoting the cumulative comment count at each timestamp for a given video. Using the time sequence vector $TS$, we train a stacked GRU network. This model learns to predict the next $l$ values for $d$ of the input variables such that $1 \leq d \leq m$. We take $m$ units in the input layer and $d \times l$ units in the output layer. The hidden layer GRU units are fully connected through recurrent connections. We stack the GRU layers in a manner that every unit in the lower GRU hidden layer is fully connected with every unit in the hidden layer above it through feed-forward connections.

We utilize the predicted values for computing the prediction error distribution using which we detect the unusual comment activities (peaks). The error vector $e^{(t)}$ is defined as $e^{(t)} = [e^{(t)}_{11}, \dots, \allowbreak e^{(t)}_{1l}, \dots, e^{(t)}_{d1}, \dots, e^{(t)}_{dl}]$, where $e^{(t)}_{ij}$ is computed by taking the difference between the actual value of $c^{(t)}_i$ and its predicted value at time $ t - j$. We fit a multivariate Gaussian distribution $\mathcal{N}(\mu,\Sigma)$ using the error vectors, where $\mu$ and $\Sigma$ are estimated using Maximum Likelihood Estimation. 

To label the observation as either collusive or non-collusive, we calculate the anomaly score $a(c)$, which is defined as the Mahalanobis distance between the computed error vector $e^{(t)}$ and the distribution $\mathcal{N}$.
\begin{equation}\label{eq:anomaly}
    a(c) = (c - \mu)^T\Sigma\inv (c - \mu)
\end{equation}

Finally, we concatenate all the observations for a given video time sequence and feed the concatenated sequence to the \textit{find\_peaks} module of Scipy\footnote{\url{https://docs.scipy.org/doc/scipy/reference/generated/scipy.signal.find_peaks.html}} to retrieve the peak width and peak height. Using the same data, we get two useful features that we pass into the collusive video detector. 

\textbf{{(i) Peak count ($\phi$):}} It is the count of the of peaks detected for a video.

\textbf{{(ii) Average peak area ($\omega$):}} The average peak area is calculated as the average of the overall area covered by each peak. After exploratory data analysis, we observe a very clear distinction between the distribution of peak width and height of collusive and other random videos. The width and height of each peak is calculated using the \textit{peak\_width} module of Scipy\footnote{\url{https://docs.scipy.org/doc/scipy/reference/generated/scipy.signal.peak_widths.html}}. 

\begin{figure}[!t]
    \centering
    \subfloat[]{{\includegraphics[width=4.8cm]{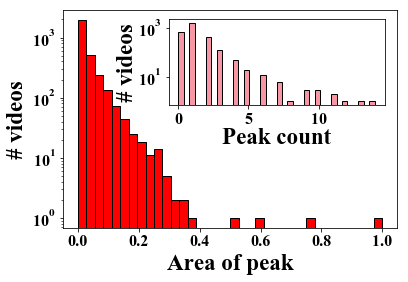} }}%
    \subfloat[]{{\includegraphics[width=4.2cm]{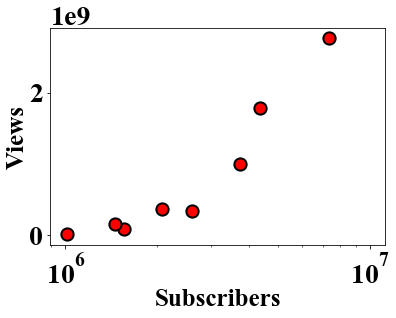} }}%
    \caption{(a) Exploratory analysis of the features used for anomaly detector in case of collusive videos. (b) Linear pattern between subscribers and views of YouTube channels i.e., channels with more subscriptions tend to lead to more views.}%
    \label{fig:analysis_anomaly}%
\end{figure}

Figs. \ref{fig:analysis_anomaly}(a) shows the peak count and average peak area for collusive videos. 


The final anomaly feature extractor \textit{$v_a$} has the following form:
\begin{equation}
    v_a = (\phi,\omega)
\end{equation}

\paragraph{Training details:}
As the idea here is to detect unusual commenting behaviour, we train GRU on \genuinevideoset~(more details on \genuinevideoset~can be found in Section \ref{sec:video_detector}) which is a set of random videos. Once trained, the model will not be able to generate patterns similar to collusive videos resulting in higher valued error vector or in simple terms -- an anomalous activity. In our experiments, we noticed that LSTM cell tends to generate multiple peaks for single activity while GRU was able to mitigate this issue well. For the same, we choose to employ GRU cells in place of LSTM cells  \cite{malhotra2015long}. Moreover, we could also reduce the model training time substantially using the GRU cells in place of the LSTM cells.

\subsubsection{Comment feature extractor}
YouTube users generally comment on videos if either they like/dislike the video or the domain of interest is the same. On the other hand, the collusive users are majorly interested in uplifting their credits just by posting a large number of comments. We observe that they tend to borrow the content from the recently posted comments, make marginal changes and post the same. We hypothesize that the comments posted during artificial boosting should have high textual similarity between the collusive comments. 

To this end, we measure the similarity of closely related comments posted during the boosting timeline.
We extract  full comment texts of all the videos using the YouTube API. The data collection strategy is detailed in Section \ref{sec:data_collection}.

\textbf{Window selection strategy}: Computing a similarity score by utilizing all the previously posted comments for a given query comment might seem intuitive but is a computationally heavy task. For the same, we first retrieve the comments posted during peak time. Moreover, we propose to use a fixed-size ($w$) moving window and roll it over the set of retrieved comments. We define the last comment in each set as query comment ($q_i$) and the other comments as the window comments ($C_i$). The best performing window size ($w$) from the experiments was found to be of size $10$.

\textbf{Comment similarity score ($\eta$)}: To calculate comment similarity score, we first encode both the query comment $q_i$ and the window comments $C_i$ using Universal Sentence Encoder (USE) \cite{cer2018universal}. As we know that collusive users can belong to different geographical places and may not use the same language, we chose to work with the multilingual USE model\footnote{\url{https://tfhub.dev/google/universal-sentence-encoder-multilingual/1}} to retrieve comment embeddings. In the literature, USE based model has been able to achieve state-of-the-art for SemEval-2017 Task \cite{cer-etal-2017-semeval} on Semantic Textual Similarity Multilingual, and Cross-lingual Focused Evaluation.
Currently, the multilingual model supports 16 different languages and has shown strong performance in cross-lingual text retrieval. The input to the model can be variable length text in any of the supported languages and the output is a 512-dimensional vector. We transform each query comment into a fix-length embedding vector $g(q_i)$ and the window comments into set of vectors $g(C_i)$ using USE.

It has been observed that the same video was posted multiple times for the collusive comments. It may also happen that the collusive user ran out of credit points due to which the collusive video is no longer shown on the blackmarket website until the user again starts earning some credit points. In such cases, our peak detection strategy is likely to detect multiple peaks ($p$) for the videos ($V$). Moreover, a peak can have a large number of comments. We define the comment similarity score ($\eta$) for a given video as follows:
\begin{equation}
    \eta = \frac{\sum_{i=1}^{P} P_{s(i)}}{P}
\end{equation}
\begin{equation}
    P_{s(i)} = \frac{\sum_{j=1}^{W} \max_{j: w_j \in P} ( g(q_j) \cdot g(C_j)^T )}{W}
\end{equation}
where $P_{s(i)}$ denotes comment similarity score for $i$th peak, \textit{$P$} denotes the number of peaks, \textit{$W$} denotes the number of windows, and $g(q_j)$ and $g(C_j)$ denote the $j$th query embedding and $j$th window embedding respectively. We choose the maximum of the dot product of query embedding and the sentence embedding under the assumption that one comment is derived from only one other comment.

We use the derived score $\eta$ along with the total comment count ($t_c$) to create  \textit{$v_c$}. The final comment feature extractor \textit{$v_c$} has the following form:
\begin{equation}
    v_c = (\eta, t_c)
\end{equation}

We then concatenate the learned representations from the metadata feature extractor \textit{$v_e$}, anomaly detector \textit{$v_a$} and comment feature extractor \textit{$v_c$} to form the video feature representation denoted as $v = v_e \oplus v_a \oplus v_c$.

\subsubsection{Collusive video detector} \label{sec:video_detector}

\begin{figure*}[!t]
    \centering
    {\includegraphics[width=\textwidth]{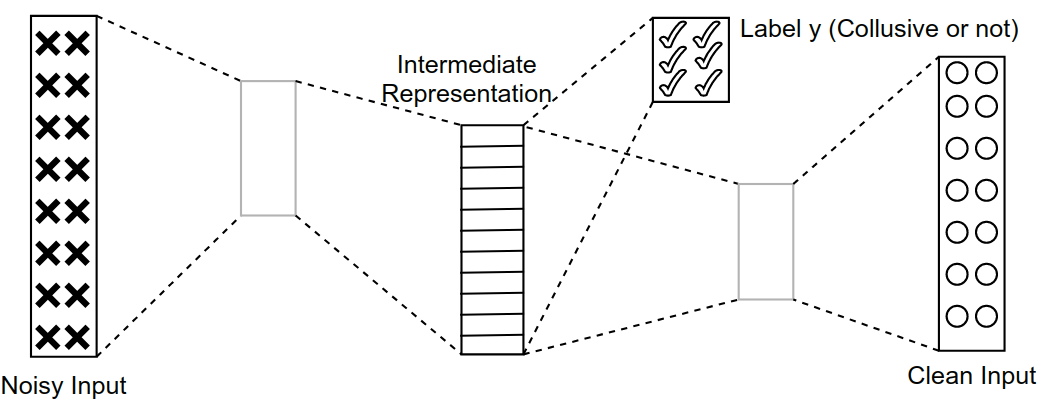} }
    \caption{The architecture of Denoising Autoencoder Classifier.}%
    \label{fig:arch2}%
\end{figure*}

Here we introduce the collusive video detector, as shown in Fig. \ref{fig:arch2}. The primary purpose of the detector is to learn the distribution from the collusive data and identify similar ones. Although this task looks very similar to the tasks mentioned in Sections \ref{sec:collusive_like_detection} and \ref{sec:collusive_channel_detection}, it actually uses a different kind of features by incorporating the temporal representation of the comments. To investigate this, we manually collected a small set of YouTube channels that are not posted on YouLikeHits. We denote this set of videos by \genuinevideoset. Here we only considered collecting the other set of videos (\genuinevideoset) from the top $5000$ channels because YouTube channels with more subscriptions tend to lead to more views, thus resembling  the collusive channels. Although it is not possible to find out view-timestamp from the YouTube API for any given video, we can still expect the overall behavior of \genuinevideoset\ to be similar to Figure \ref{fig:analysis_anomaly}(b) as a user-driven blackmarket service is analogous to a channel in a few ways such as both benefit from more subscribers, the subscribers are loyal and tend to watch/like/comment on the videos posted. We selected a playlist from each channel at random. From each playlist, we randomly selected \textit{k} (where, \textit{k}=$3$ in our case) videos. As the overall data was biased towards some specific genres, we removed some videos at random to maintain generality in data. The presence of such similar patterns in both the classes increases the difficulty of the classification task even further but will make the classifier more relevant for real-world collusion detection. Here  \genuinevideoset\ contains videos having similar comment growth patterns and thus can be considered a noisy/adversarial set in terms of comment growth patterns. Note that it does not represent the entire YouTube population. For the same, we report the true positive rate (TPR) of the model with respect to the set of collusive videos.  

Training a fully connected network that directly optimizes only the supervised objective by gradient descent also does not work very well in such cases. What works better is to initially use a local unsupervised criterion to pre-train each layer, with the goal of learning to produce a useful higher-level representation from the lower-level representation output by the previous layer. From this initial point, gradient descent on the supervised objective leads to better solutions in terms of generalization performance \cite{vincent2010stacked,ZHANG201892}.

We propose to use Denoising Autoencoder Classifier (DAC), which uses autoencoders, a deep unsupervised learning method that improves the generalization of supervised learning on limited labeled data. In particular, we investigate a case where the input signal is noisy and we explore the multi-task learning with two tasks: a classification output, coupled with a reconstruction output of a Denoising Autoencoder. To establish the effectiveness of this approach, we also present a comparison of this approach against different baselines, including a multi-layer perceptron (MLP) model as a replacement of the Denoising Autoencoder (more details in Table \ref{table:performance}).

Throughout this section, we define the input space as \textit{X}, input data point \textit{x} $\in$ \textit{X}, label space as \textit{Y}, output label \textit{y} $\in$ \textit{Y}, intermediate decoded space \textit{V}, intermediate encoding \textit{z} $\in$ \textit{Z}, noisy or corrupted input data point \textit{$\hat{x}$} $\in$ \textit{X}.

At a high level, DAC consists of two structural components: the denoising autoencoder, and the dense layer as the classifier. Denoising autoencoder has a very similar structure to the autoencoders, except rather than reducing only the reconstruction loss over the input data, it maps noisy inputs to original clean inputs. Autoencoders consist of two components: an encoder $(\tau: X \implies Z)$ and a decoder $(\psi: Z \implies X)$ where $\tau$ and $\psi$ are inversely related. They first map an input $x \in X \in \mathbb{R}^d$ to an intermediate representation $z \in Z \in \mathbb{R}^{d^{\prime}}$ of reduced dimensionality $(d^{\prime} < d)$ via $\tau$. This intermediate representation is then reconstructed back to a point $x^{\prime}$ in the input space, and the difference between $x$ and $x^{\prime}$ is measured using a loss function $L$.

On the contrary, in case of denoising autoencoder, a noisy representation of $x$, \textit{$\hat{x}$}, is fed to the network yielding \textit{$\hat{x}^{\prime}$}, and then  \textit{$\hat{x}^{\prime}$} is compared to the original clean input data point $x$. The classifier utilizes the intermediate representation $z$ of the denoising autoencoder. After encoding noisy input, the classification task is added that contributes to the encoder weights independently from the decoder. This task consists of a fully-connected layer that predicts the label $y^{\prime} \in Y$. Finally, both the classification loss and reconstruction loss propagate backward and contribute to encoder weights.

\textbf{Training details:}
We consider a fully-connected autoencoder with two hidden layers containing 128 units each for all the experiments. We use mean squared error as the loss function difference between $x$ and \textit{$\hat{x}^{\prime}$}. To generate corruption in training data, we randomly corrupt the input data by $\pm10\%$ of its original value.
We use the entire unlabeled training set, i.e., $1384$ videos -- $756$ collusive and $628$ random videos, to train the denoising autoencoder for 25 epochs and minimize mean squared error (MSE) from the reconstruction output. During this stage, we do not update the weights for edges between the encoded layer and classification output. The denoising autoencoder classifier is then trained to minimize the categorical cross-entropy loss on the labeled training samples for 150 epochs. Note that we choose the optimal value of each parameter based on the hyperparameter search. We implement our model using Python 3.8.3 and TensorFlow 2.3.1. The model has a total of $18k$ trainable parameters and is less than $100KB$ in size.





%
\section{Experimental results}\label{sec:result}
\subsection{Baselines for detecting videos submitted for collusive comments}
We validate the performance and robustness of \collate~by comparing with seven baselines. Since there is no existing research dealing with the same problem present in this paper, we develop a few of our own baselines and adopt some of the existing studies to our problem setting.
 We report the performance of all the competing methods along with the detailed analysis of the performance and robustness. 
\begin{itemize}
    \item \textbf{Extractor models:}
\collate\ uses the combination of metadata, anomaly and comment feature extractors. Thus, we propose two baselines: (i) considering only one extractor at a time ($B_1$), and (ii) considering two extractors at a time ($B_2$). This also provides the ablation study of our method.

\item \textbf{Variant of proposed model ($B_3$):}
As shown in Fig. \ref{fig:architecture}, comment feature extractor and anomaly feature extractor can be modified to obtain a variant of \collate. We modify both the components to obtain different variants of the proposed model as described below:

\textit{LSTM-based Anomaly Extractor}: We replace LSTM units in place of GRU for detecting the anomalous comments. We observe that this variant tends to generate multiple peaks for the videos with only one-time collusive comments. This change not only adds the noisy comments but also deteriorates the overall performance significantly as the comment feature extractor is dependent on this stage.

\textit{WMD-based Comment Similarity Score}:
Word Mover's Distance (WMD) \cite{kusner2015word} is a popular metric for calculating semantic similarity at a document level. We utilize standard pre-trained Word2Vec embeddings from Gensim library\footnote{We tried with glove-twitter-200 and word2vec-google-news-300 embeddings and found the latter to perform better.} to compute the Word Mover's Distance $d_{ij}$. We transform the distance $d_{ij}$ into a similarity score $s_{ij}$ as below:
\begin{equation}
    s_{ij} = \frac{1}{1 + d_{ij}}
\end{equation}
where $i$ denotes the comment index, and $j$ denotes the window index. Rest of the details for computing $\eta$ remains the same as mentioned in Section \ref{sec:collusive_comment_detection}. 

There are two major differences between this model and \collate\ --  (i) word-level vs. sentence-level embeddings, and (ii) monolingual vs. multilingual support. We discuss the impact of these differences through our experiment results in the next section.

\item \textbf{DetectPV ($B_4$):} We use the method proposed by Bulakh et al.\cite{bulakh2014identifying} as our fourth baseline. It uses a supervised learning approach to identify fraudulently promoted videos by extracting features from the video metadata. We wanted to check if video metadata-based fraudulent video detection techniques would be useful to detect collusive videos on YouTube.  

\item \textbf{ARIMA ($B_5$):} Since \collate\ uses temporal information of comments posted on videos to detect peaks and their associated properties, one may argue that a time series based anomaly detection method may be able to detect collusive videos. To this end, we consider the method proposed by \cite{box1970distribution} as another baseline. ARIMA is an auto-regressive integrated moving average model that uses the combination of auto-regression and moving average to detect anomalies on time-series data.

\item \textbf{CNN ($B_6$):} Feature interaction-based prediction models such as convolutional neural network (CNN) models have been used extensively for various image processing tasks \cite{sindagi2018survey,tu2019survey}. In our case, we use 1-D CNN layer, followed by dense layers to form a fully-connected neural network. The model learns to extract and map features from the data to the output labels.

\item \textbf{WND ($B_7$):} We use the Wide \& Deep Factorization (WND) method proposed in Guo et al.\cite{guo2017deepfm} as another baseline. The model consists of a wide part (to memorize the past behavior) and deep part (to embed into lower dimension) and utilizes deep neural network and factorization machines to address the interactions among the features.

\end{itemize}

\subsection{Prediction results} \label{sec:prediction_results}
\textbf{Tasks 1 and 2: Identify the videos (channels) submitted to blackmarket services for collusive likes (subscriptions).} \\
Due to the restriction of the YouTube Data API, we are unable to access the timestamp of like/dislike activity.  This resists us to create any time-centric features for collusive like appraisals. Similarly, we are unable to get detailed information about YouTube channels from the API. For our prediction task, we use one-class classification models with the features mentioned in Sections \ref{sec:collusive_like_detection} and   \ref{sec:collusive_channel_detection}. The models are trained on one class, which in our case is the collusive class. Note that we only report the \textit{true positive rate} (TPR) for our prediction task as we are only interested in the proportion of actual positives that are correctly identified by the model. We perform the prediction task with the following state-of-the-art one-class classifiers: one-class SVM \cite{scholkopf2001estimating}, isolation forests \cite{liu2008isolation}, minimum covariance determinant \cite{hubert2018minimum}, and local outlier factor \cite{breunig2000lof}. All the scores are reported after 5-fold cross validation. Note that the test set remains same across all the competing methods.

\begin{figure}[!t]
    \centering
    \subfloat[]{{\includegraphics[width=6.7cm]{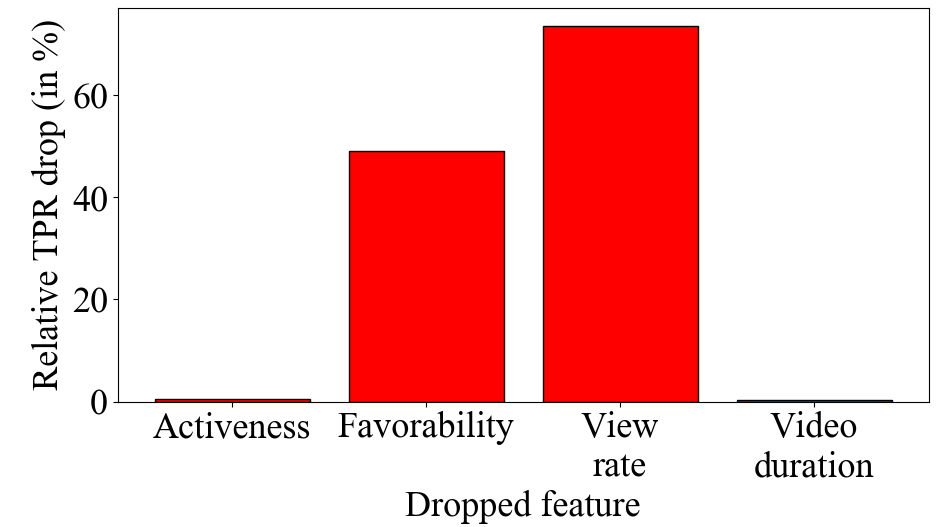} }}%
    \subfloat[]{{\includegraphics[width=5.8cm]{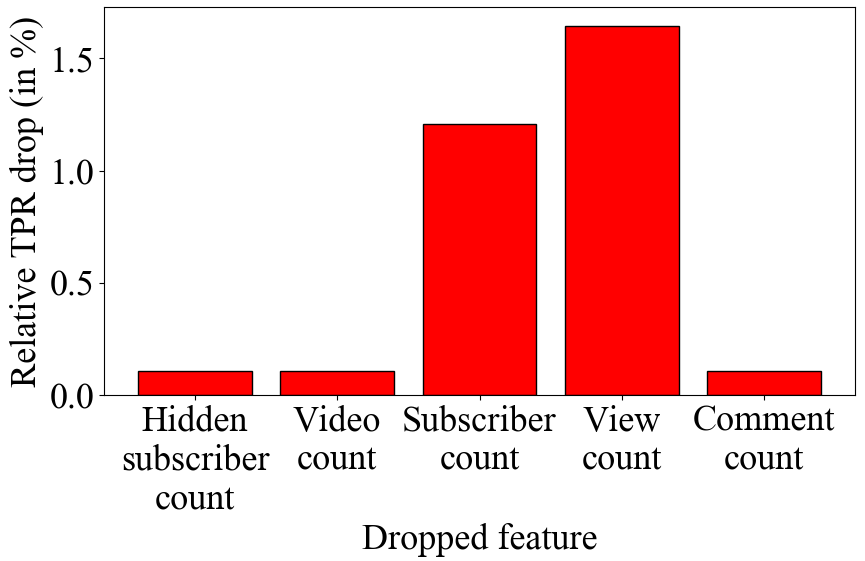} }}%
    \caption{Feature importance considering (n-1) features at a time, i.e., dropping each feature in isolation (a) for Task 1 and (b) Task 2.}%
    \label{fig:feature_imp}%
\end{figure}

Table \ref{table:performance_model_12} reports the performance of the one-class classification models for detecting videos submitted for collusive likes (Task 1) and channels submitted for collusive subscriptions (Task 2).
In the former task, we observe the best accuracy of the model (with true positive rate of $0.911$) with one-class SVM.  To analyze the influence of each feature, we perform experiments, taking $(n-1)$ features at a time, i.e., dropping each feature in isolation. The most important feature turns out to be \textit{view rate} based on quantified relative importance. In the latter task, we once again observe that one-class SVM performs the best (with true positive rate of $0.910$).  \textit{View count} turns out to be the best feature for this task. Fig. \ref{fig:feature_imp} shows the feature importance for Task 1 and Task 2 with the relative TPR drop percentage when we drop each feature in isolation.   \\

\begin{table}[!t]
\centering
\caption{Performance of one-class classification models on two tasks. Task 1: videos submitted for collusive likes; Task 2:  channels submitted for collusive subscriptions.}\label{table:performance_model_12}

\begin{tabular}{|c|c|c|}
\hline
\textbf{Model} & \textbf{Task} & \textbf{True positive rate} \\\hline
One-class SVM & Task 1 & \textbf{0.911} \\
 & Task 2 & \textbf{0.910} \\ \hline
Isolation forests & Task 1 & $0.902$ \\
 & Task 2 & $0.906$ \\\hline
Minimum covariance determinant & Task 1 & $0.900$ \\
 & Task 2 & $0.901$ \\\hline
Local outlier factor & Task 1 & $0.899$ \\
 & Task 2 & $0.894$ \\\hline

\end{tabular}
\end{table}

\noindent \textbf{Task 3: Identify the videos submitted to blackmarket services for collusive comments.}\\
Our third task is to determine whether a given video was submitted to blackmarket services for collusive comments. We use the approach explained in Section \ref{sec:collusive_comment_detection} to detect if a video is collusive. Table \ref{table:performance} shows that DAC consistently achieves better accuracy compared to its baselines. In Table \ref{table:performance}, we mark the best performing extractor for each of the baselines, $B_1$, $B_2$ and $B_3$ in bold. 

Here, CFE-USE denotes the usage of Universal Sentence Encoder while CFE-WMD denotes the usage of Word Mover's distance for calculating sentence similarity. CFE-USE (\textit{comment similarity} and \textit{comment count}) with DAC is individually the best performing extractor for baseline $B_1$. Although we expect AFE to perform better than at least MFE, due to the nature of the non-collusive video data, it does not. These results support our argument about the higher similarity in the distribution for both collusive and non-collusive datasets. For the same, we take into account multiple extractors that not only improvise the performance but also increase the robustness. $B_2$ (MFE + CFE-USE) with DAC  outperforms other permutations as it captures the most useful representation for classification.   

$B_3$ (\textit{CFE-WMD}) seems to perform better than a few baselines but is unable to outperform \collate. This is expected because WMD-based similarity score calculation depends on word-level embeddings rather than sentence-level embeddings. Moreover, we also find that the feature importance is evenly distributed in \collate~compared to $B_3$. \citet{kolcz2009feature} also suggested that the weight vector of a robust classifier should be distributed as evenly as possible. Finally, we report the baseline performance of $B_4$ (DetectPV), $B_5$ (ARIMA), $B_6$ (CNN) and $B_7$ (WND). We observe that the overall performance of \collate~is better than the baselines.

\begin{table}[!htbp]
\centering
\caption{Performance comparison of \collate\ with baselines for detecting videos submitted for collusive comments.}\label{table:performance}

\begin{tabular}{|l||c|c|c|c|}
\hline

\bf{Method} & \bf{TPR} & \bf{FPR} & \bf{Accuracy} & \bf{AUC}\\ \hline
$B_1$ (AFE) & $0.715$ & 0.455 & 0.638 & 0.629\\
$B_1$ (CFE-USE) (MLP) & $0.789$ & 0.410 & 0.718 & 0.689\\
$B_1$ (MFE) & $0.825$ & 0.444 & 0.702 & 0.690\\
$B_1$ \textbf{(CFE-USE) (DAC)} & \textbf{0.866} & 0.157 & 0.845 & 0.834\\
\hline
$B_2$ (MFE $+$ CFE-USE) (MLP) & $0.792$ & 0.395 & 0.704 & 0.698\\
$B_2$ (MFE $+$ AFE) & $0.795$ & 0.412 & 0.701 & 0.692\\
$B_2$ (AFE $+$ CFE-USE) (DAC) & 0.875 & 0.177 & 0.839 & 0.848\\
$B_2$ \textbf{(MFE} $+$ \textbf{CFE-USE) (DAC)} & \textbf{$0.881$} & 0.179 & 0.853 & 0.851\\ \hline
$B_3$ \textbf{(AFE + MFE + CFE-WMD) (DAC)} & \textbf{0.866}  & 0.182 & 0.844 & 0.841\\ \hline 
$B_4$ (DetectPV) & $0.829$  & $0.232$ & $0.802$ & 0.798\\
\hline 
$B_5$ (ARIMA) & $0.768$  & $0.429$ & 0.677& 0.669\\
\hline
$B_6$ (CNN) & $0.679$  & $0.491$& 0.609 & 0.594\\ \hline
$B_7$ (WND) & $0.890$  & 0.250 & 0.825 & 0.820\\
\hline \hline
\textbf{\collate} & \textbf{0.905} & 0.194 & \textbf{0.860} & \textbf{0.855} \\ \hline

\end{tabular}
\end{table}

\section{Interesting Observations} \label{sec:implications}
In this section, we detail the important implications of collusive entity detection task. To this end, we consider only the top $10\%$ of the collusive entities detected (true positive) using our models to present our analysis. For tasks 1 and 2, the top $10\%$ is generated using the scoring function\footnote{\url{https://tinyurl.com/ve5yvr47}} of our best-performing one-classification model. For task 3, the top $10\%$ is generated using the softmax values present in the last layer of \collate. We use the term \textit{highly collusive} to refer to the collusive entities present in the top $10\%$ in each of the cases.

\begin{observation}
[{\bf Highly collusive videos have high video ratings}]
\label{obs_1}
We define rating as the ratio of likes to the sum of likes and dislikes gained by the video -- the video will have rating $1$ if there are no dislikes. We observe that highly collusive videos have very high ratings with an average rating of $0.931$. This shows how blackmarket services have been able to gain collusive appraisals in an effective way. 
\end{observation}

\begin{observation}
[{\bf Highly collusive videos have very short video duration}]
\label{obs_2}
We note that highly collusive videos have a very short duration, with average video length of only \textasciitilde$4$ minutes. As confirmed by YouTube\footnote{\url{https://youtube-creators.googleblog.com/2012/08/youtube-now-why-we-focus-on-watch-time.html}}, it itself promotes videos that keep people on YouTube for a long period of time. Thus, videos with short duration do not get the proper audience naturally, thereby making the authors choose blackmarket services to gain artificial appraisals in a quicker way.
\end{observation}

\begin{observation}
[{\bf Gaining collusive likes does not guarantee to gain collusive comments}]
\label{obs_3}
We observe that the videos submitted for collusive likes do not have many comments in them, with the average number of comments being only $12.4$. This also corroborates with the fact that we do not have any intersection between the sets $V_l$ and $V_c$. We can ensure that collusive like requests and collusive comment requests are two completely independent activities on the blackmarket platforms.
\end{observation}

\begin{observation}
[{\bf Highly collusive channels are popular channels with a large number of videos}]
\label{obs_4}
We note that highly collusive channels are popular YouTube channels with average subscribers count of $39,477$, average view count of $8,762,188$, and average videos count of $111$. The reason  is that getting new subscribers for those channels is an extremely difficult task, which makes them choose artificial ways of gaining new subscribers by means of the blackmarket services. 
\end{observation}

\begin{observation}
[{\bf Highly collusive videos have a moderate inter-arrival rate of comments}]
\label{obs_5}
We study the inter-arrival rate of comments in the highly collusive videos. Surprisingly, we observe a mean inter-arrival rate of \textasciitilde$5$ hours for each comment. The possible reason behind such high value of inter-arrival rate is due to the expiration of credits for the collusive comments in the blackmarket services.
\end{observation}

\begin{observation}
[{\bf Highly collusive videos have a moderate inter-arrival rate of comments}]
\label{obs_5}
We study the inter-arrival rate of comments in the highly collusive videos. Surprisingly, we observe a mean inter-arrival rate of \textasciitilde$5$ hours for each comment. The possible reason behind such high value of inter-arrival rate is due to the expiration of credits for the collusive comments in the blackmarket services.
\end{observation}

\begin{observation}

({\bf `People \& Blogs' is the most popular genre for highly collusive channels})
\label{obs_5}
We observe `People \& Blogs' to be the most popular genre ($33.53\%$) for highly collusive channels. This genre usually contains YouTube bloggers who upload original content, and share it all with friends, family, and the world on YouTube. This is due to the reason that blogging in YouTube is currently considered as the most profitable way to earn money\footnote{\url{https://medium.com/@KeywordsHeaven/blogging-vs-youtube-72803bb3dacf}}. Thus, to gain quick popularity and recognition in the community, these users join blackmarket services. The second most popular genre  for highly collusive channels is `Music' ($22.60\%$).
\end{observation}

\begin{observation}
({\bf Highly collusive videos are family-friendly and non-paid unlisted videos})
\label{obs_5}
We observe that highly collusive videos are family-friendly and non-paid videos. The possible reason behind not submitting violent and mature YouTube videos to blackmarket services may be due to the reason that YouTube does not allow age-restricted videos to be monetized. The possible reason behind not submitting paid videos to blackmarket services is that the monetization of YouTube videos is determined by the level of engagement (likes/comments) a video generates.
\end{observation}

\begin{figure}[!t]
    \centering
    \subfloat[]{{\frame{\includegraphics[width=4.1cm]{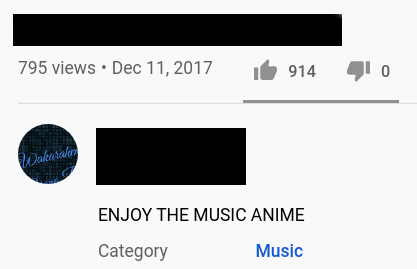}} }}%
    \subfloat[]{{\frame{\includegraphics[width=4.7cm]{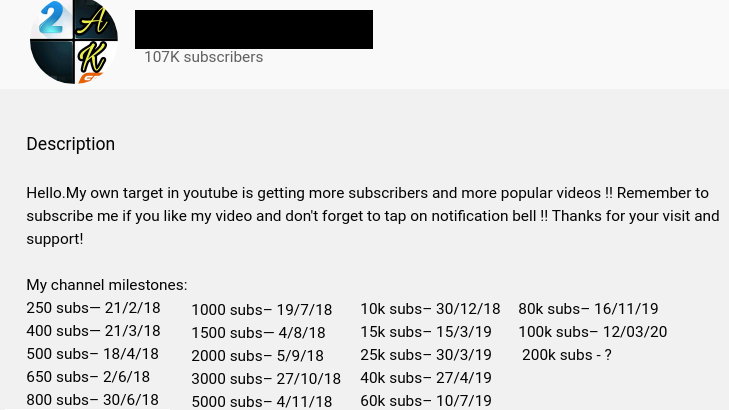} }}}%
    \hspace{0.26em}%
    \subfloat[]{{\frame{\includegraphics[width=4.5cm]{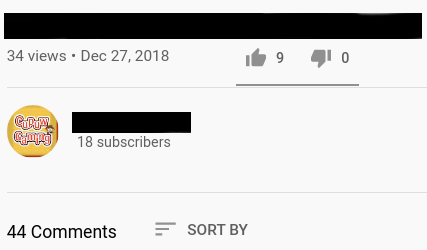}} }}%
    \caption{Example of (a) collusive video (for likes), (b) collusive channel (for subscriptions), and (c) collusive video (for comments) detected by our models. Sensitive information are blurred.}%
    \label{fig:model_examples}%
\end{figure}

Fig. \ref{fig:model_examples} shows the snapshot of some collusive entities detected using our models. Fig. \ref{fig:model_examples}(a) shows a collusive YouTube video where the number of views is much lesser than the number of likes. This clearly gives us the indication that the YouTuber is buying artificial likes from the blackmarket websites. Fig. \ref{fig:model_examples}(b) shows a collusive YouTube channel where the description clearly states that the author is looking for quick popularity. The YouTube is maintaining a timeline of the increase in the subscriber count of the channel. Also, the channel description contains promotional keywords such as `gaining more subscribers'. Fig. \ref{fig:model_examples}(c) shows a collusive YouTube video where the video has more number of comments than the views and likes. This also gives us the indication that the YouTuber is buying artificial comments from the blackmarket websites.

\section{Conclusion and Future Work}\label{sec:conclusion}
With the increase in the involvement of content creators and content consumers on YouTube, social growth has become the most important metric for popularity. To achieve rapid  social growth, a large number of content creators go against the stream by registering their videos/channels to blackmarket services for collusive appraisals. Although many studies have been carried out to detect fraud and fake activities on multiple online media platforms, identification of collusive entities  remains a relatively important unexplored area of research. The major contributions of this work are manifold. (i) We collected a large dataset of YouTube videos submitted to blackmarkets for collusive likes, comments and subscriptions. \textit{To the best of our knowledge, this is the first dataset of this kind}. (ii) We analyzed the collusive videos from two perspectives: propagation dynamics and video metadata. The YouTube channels are analyzed based on their location, video metadata and network properties. (iii) In order to detect videos submitted in blackmarket services for collusive likes and channels submitted for collusive subscriptions, we utilize one-class classification models trained only on the collusive data. The SVM-based model achieves $0.911$ TPR using video metadata features and $0.910$ TPR using channel features to detect videos and channels submitted to blackmarket services for collusive likes and
subscriptions respectively. (iv) We then proposed \collate, a system that combines  three feature extractors (metadata, anomaly and comment) to learn representations of a video. \collate\ makes use of extractors effectively for identifying whether a video is registered in blackmarket services for collusive comment appraisals. Extensive experiments on our dataset show that \collate\ is effective in detecting collusive entities with $0.905$ TPR.
(v) As a final contribution, we show the important implications of the collusive entity detection task. We expect this research to push further studies in online media outlets to explore the dynamics of collusive behavior. 

We believe our proposed methodologies can be  helpful for a variety of tasks in similar research. For the first two models, i.e., identifying videos (channels) submitted to blackmarket services for collusive likes (subscriptions), it can only be adopted to video-sharing platforms as the features used in the models are generated from video (channel) metadata present on YouTube. In case of \collate, the \textit{anomaly feature extractor} and \textit{comment feature extractor} can be adopted by any video or non-video sharing platforms where a user posts texts on some entities of that platform. E.g., in case of Twitter to detect users who submit tweets to blackmarket services for collusive retweet appraisals, the \textit{anomaly feature extractor} can utilize the retweet time-series data to detect sudden changes in the retweeting activity;  \textit{comment feature extractor} can utilize the text of the comments posted during collusion and compute a similarity score between the comments. However, we have not performed this experiment in the current paper since the major aim of this paper is to provide a detection framework for collusive YouTube entities, not developing a framework for detecting collusive entities on Twitter which we explored in our previous studies \cite{dutta2018retweet,dutta2019blackmarket,chetan2019corerank,arora2019multitask,arora2020analyzing}.

Despite encouraging results, collusive entities detection still remains a challenging problem with many open research questions. In the future, we are interested in exploring the following avenues. First, we plan to take into account the sentiment of the comments by considering the average comment sentiment during collusion for detecting the collusive activity. Second, we wish to study the interdependency of the collusive videos and collusive users that can help in identifying the core users of the blackmarket services. Third, we intend to detect collusive entities at an inter-platform level as well. Fourth, our final goal will be to design a web-based scalable collusive entity detection system for online video sharing platforms.

\section*{Acknowledgement}
The authors would like to thank the support of ECR/2017/001691
(SERB), Ramanujan Fellowship, and ihub-Anubhuti-iiitd Foundation (set up under the NM-ICPS
scheme of the Department of Science and Technology, India).

\bibliographystyle{ACM-Reference-Format}
\bibliography{bibliography}

\end{document}